\documentclass[prc,twocolumn,showpacs]{revtex4}

\usepackage{graphicx}
\usepackage{dcolumn}
\usepackage{bm}

\begin{document}

\title{$\beta$-decay properties of neutron-rich Ca, Ti, and Cr isotopes}

\author{P. Sarriguren$^{1}$}
\email{p.sarriguren@csic.es}
\author{A. Algora$^{2,3}$}
\author{G. Kiss$^{3,4}$}
\affiliation{
$^{1}$ Instituto de Estructura de la Materia, IEM-CSIC, Serrano
123, E-28006 Madrid, Spain \\
$^{2}$ Instituto de F\'\i sica Corpuscular, CSIC-Universitat de Valencia, E-46071 
Valencia, Spain \\
$^{3}$ Institute of Nuclear Research of the Hungarian Academy of Sciences, 
4026 Debrecen, Hungary \\
$^{4}$ RIKEN Nishina Center, Wako Saitama 351-0198, Japan \\
}


\date{\today}

\begin{abstract}

$\beta$-decay properties of neutron-rich Ca, Ti, and Cr isotopes are studied 
within a deformed proton-neutron quasiparticle random-phase approximation. 
The underlying mean field is described self-consistently from deformed Skyrme 
Hartree-Fock calculations with pairing correlations. Residual spin-isospin
interactions in the particle-hole and particle-particle channels are also 
included in the formalism. 
The energy distributions of the Gamow-Teller strength, the $\beta$-decay 
feedings, the $\beta$-decay half-lives, and the $\beta$-delayed neutron 
emission probabilities are discussed and compared with other theoretical 
results, as well as with the available experimental information.
The evolution of these nuclear $\beta$-decay properties is investigated in 
isotopic chains in a search for structural changes. A reliable estimate of 
the $\beta$-decay properties in this mass region is a valuable information 
for evaluating decay rates in astrophysical scenarios.

\end{abstract}


\maketitle

\section{Introduction}

The study of neutron-rich nuclei in the mass region determined by $Z \geq 20$ and 
$N \geq 28$ has attracted recently increasing interest from both theoretical 
and experimental sides for different reasons.
From a purely nuclear structure point of view, one of the most intriguing 
peculiarities characterizing this region is the appearance of effects compatible 
with sub-shell closures.
As a matter of fact, the existence of a sub-shell closure at $N=32$ seems to be 
well established experimentally in Ca, Ti, and Cr according to the systematics 
of the $E(2^+_i)$ energies in the isotopic evolution and from recent mass
measurements performed at ISOLDE/CERN \cite{wienholtz13}. $\beta$-decay properties, 
the identification of the first excited states, and deduced $B(E2:2^+\rightarrow 0^+)$ 
transition probabilities support that conclusion as well, and in addition indicates 
the existence of an $N=34$ shell closure, see \cite{crawford10,steppenbeck13,mantica08} 
and references therein.  In this context it is also of interest that the $N=32$ 
shell closure was recently challenged by the unexpectedly large charge radii 
measurements reported in \cite{garcia16}, but apart from the interpretation provided 
in \cite{garcia16} alternative explanations exist \cite{saperstein16}. 
It has also been argued from shell-model calculations that an eventual inversion 
of the $\nu 1f_{5/2}$ and $\nu 2p_{1/2}$ orbitals could create in this mass
region a sub-shell closure at $N=34$ \cite{honma05}. Indeed, the spectroscopy 
of high spin states in Ti isotopes suggests that these orbitals are practically 
degenerate \cite{fornal2004}. With increasing neutron number in the vicinity of 
$N=40$ a new island of inversion is predicted and deformation sets 
in \cite{lenzi10,crawford13,aoi09,gade14}.

As we move away from stability, finding magicity evidence from spectroscopic 
studies becomes more and more difficult because of the present limitations 
regarding production rates and because of the short lifetimes inherent to 
the exotic nuclei.
Thus, the study of other indirect indications based on the decay properties, 
such as $\beta$-strength distributions, $\beta$-decay half-lives $T_{1/2}$, and
$\beta$-delayed neutron emission probabilities $P_n$, might be worthwhile.
The decay patterns of such exotic nuclei represent a unique possibility to 
learn about their nuclear structure. As an example, a sudden shortening of 
half-lives, associated with magicity, was observed in Ni isotopes beyond $N=50$ 
\cite{xu2014}, but not in neighbor chains.

Whereas the $\beta$-strength distribution contains detailed information about 
the structure of nuclei involved in the decay, $T_{1/2}$ and $P_n$ are integral
quantities that characterize globally the decay.
The ideal test for a nuclear model would be to compare the calculated 
$\beta$-strength distribution with measurements, but unfortunately, 
this is a difficult experimental task. $\beta$-decay half-lives and $P_n$ 
probabilities are, on the other hand, easier to measure.
In particular, $P_n$ are very interesting quantities in neutron-rich nuclei.
As we move away from stability toward the neutron-rich side, the neutron 
separation energy $S_n$ decreases until eventually a zero value is reached at 
the drip lines. When $S_n$ becomes lower than the $Q_\beta$ energy, neutron-unbound
excited states above $S_n$ can be populated by $\beta$-decay in the daughter nucleus 
that subsequently may decay by neutron or $\gamma$-ray emission. $\beta$-delayed 
neutron emission appears as an additional decay mechanism characterized by the 
emission of neutrons after $\beta$-decay. We should not forget that $P_n$ is 
sensitive to the $\beta$-strength distribution and thus, to the nuclear structure 
itself.

The interest in neutron-rich nuclei in this mass region is not only due to their 
special nuclear structure characteristics. 
In nuclear reactor physics, $\beta$-decay half-lives and delayed neutron emission 
of the sub-products of the fission processes \cite{valencia17} are also crucial 
quantities for controlling safety in reactors. 
$\beta$-decays of nuclei in this mass 
region play also a key role in nuclear astrophysics because they take part in
the URCA process \cite{gamow45}. This is a cooling mechanism acting on white 
dwarfs, type Ia supernovae, and neutron stars, which is caused by neutrino and 
anti-neutrino emission generated in cycles of electron captures and 
$\beta^-$-decays, respectively. Nuclei in the mass region studied in this work 
are among those identified to have the highest cooling rates in the URCA 
process \cite{schatz14}.
Proper understanding of this mechanism requires knowledge of the decay 
properties of neutron-rich nuclei and the underlying nuclear physics may play 
a role comparable to the astrophysical environment to account quantitatively 
for these processes. 

Besides shell-model type calculations \cite{langanke03,zhi13,kumar16}, 
the quasiparticle random-phase approximation has proved over the years to 
be a well suited model to describe medium-mass open-shell nuclear properties 
and particularly $\beta$-decay properties within the proton-neutron
quasiparticle random-phase approximation (pnQRPA). pnQRPA calculations for 
neutron-rich nuclei have been carried out within different spherical 
formalisms based on Hartree-Fock-Bogoliubov method
\cite{engel}, on Fayans density functionals with continuum QRPA
\cite{borzov3}, and on relativistic mean field approaches 
\cite{niksic05,marketin16}.
Deformed pnQRPA formalisms based on phenomenological mean fields with separable 
residual forces \cite{moller1,moller3,homma,hir2} and with more realistic 
CD-Bonn residual forces \cite{fang13,ni14} are also available. Various 
self-consistent deformed pnQRPA calculations to describe the $\beta$-decay 
properties, either with Skyrme \cite{yoshida13,mustonen16} or Gogny 
\cite{peru14} interactions are also available in the literature. 

In Refs. \cite{sarripere,sarri14,sarri15} the decay properties of neutron-rich 
isotopes in the mass region $32<Z<46$ and $50<N<82$ were studied within a 
deformed pnQRPA based on a self-consistent Hartree-Fock (HF) mean field 
formalism with Skyrme interactions and pairing correlations in the BCS approximation. 
Residual spin-isospin interactions were also included in the particle-hole (ph)
and particle-particle (pp) channels \cite{sarri1,sarri2}. The study was extended 
in Ref. \cite{sarri_rare} to cover the decay properties of neutron-rich 
rare-earth isotopes. The reliability of the method was also tested experimentally 
with the decay properties of deformed neutron-deficient medium-mass isotopes
\cite{sarri_wp,sarri_rp}.
The purpose of this work is the study of the unstable even-even $^{50-64}$Ca, 
$^{52-66}$Ti, and $^{56-70}$Cr isotopes within a similar theoretical formalism. 
The calculations will be tested with the available experimental information on 
half-lives. Then, predictions are made for the Gamow-Teller (GT) strength 
distributions and for the half-lives and neutron emission probabilities of 
more exotic nuclei not yet measured.

These calculations are timely because the mass region addressed is at the 
borderline of present experimental capabilities for measuring half-lives 
and $\beta$-delayed neutrons. There is increasing experimental activity
focused to extend our knowledge about the decay properties of neutron-rich 
nuclei in different mass regions 
\cite{crawford10,xu2014,sorlin99,hosmer05,pereira,nishimura11,caballero16,madurga16,wu17}.
See also Ref. \cite{dillmann18} for a recent review of the experimental status 
at RIKEN on this topic. GT strength distributions of stable $f$-shell nuclei 
have been also measured with charge-exchange reactions \cite{fujita14}.

The paper is organized as follows. In Sec. II we present briefly the theoretical 
formalism needed for the calculation of the $\beta$-decay properties. In Sec. III 
we report and discuss our results for the energy curves, GT strength distributions, 
half-lives, and $\beta$-delayed neutron emission. Section IV summarizes the main 
conclusions.

\section{Theoretical Formalism}
\label{sec2}

A brief summary of the theoretical formalism used in this paper to describe 
the $\beta$-decay properties in neutron-rich isotopes is presented here. 
Further details can be found elsewhere \cite{sarri1,sarri2}.

We start from a self-consistent calculation of the mean field in terms of a 
deformed Hartree-Fock calculation with Skyrme interactions and pairing 
correlations in the 
BCS approximation. The Skyrme interaction SLy4 \cite{sly4} is selected for the 
calculations because of its ability to account successfully for a large variety 
of nuclear properties all along the nuclear chart \cite{bender08,stoitsov}. 
Single-particle energies, wave functions, and occupation amplitudes are generated 
in this way. The solution of the HF equations is found by using the formalism 
developed in Ref. \cite{vautherin}, assuming time reversal and axial symmetry. 
The single-particle wave functions are expanded into the eigenstates of a 
harmonic oscillator with axial symmetry in cylindrical coordinates, using 
twelve major shells. The pairing gap energies for protons and neutrons in the 
BCS approximation are determined phenomenologically from the experimental 
odd-even mass differences \cite{audi12}. We also perform constrained HF 
calculations to construct potential-energy curves (PECs), where the HF energy 
is minimized under the constraint of keeping fixed the nuclear quadrupole 
deformation.

In the next step, the $\beta$-decay strengths are calculated for the 
equilibrium shapes of each nucleus, that is, for the  minima obtained in the 
PECs. Since decays connecting different shapes are disfavored, similar 
shapes are assumed for the ground state of the parent nucleus and 
for all populated states in the daughter nucleus \cite{moller1,homma,boillos}.

To describe GT transitions, we add to the mean field separable spin-isospin 
residual interactions in the ph and pp channels, which are treated in a deformed 
pnQRPA \cite{moller1,moller3,homma,hir2,sarri1,sarri2}. 
According to previous calculations within this formalism
\cite{sarripere,sarri14,sarri15,sarri_rare}, we use the values
$\chi ^{ph}_{GT}=0.15$ MeV and $\kappa ^{pp}_{GT} = 0.03$ MeV for 
the coupling strengths of the residual interaction in the ph and pp channels, 
respectively.

The GT transition amplitudes in the intrinsic frame connecting the 
ground state $| 0^+\rangle $ of an even-even nucleus to one phonon 
states in the daughter nucleus $|\omega_K \rangle \, (K=0,1) $ are 
found to be

\begin{equation}
\left\langle \omega _K | \sigma _K t^{-} | 0 \right\rangle =
\sum_{\pi\nu}\left( q_{\pi\nu}X_{\pi
\nu}^{\omega _{K}}+ \tilde{q}_{\pi\nu}Y_{\pi\nu}^{\omega _{K}}
\right) 
 \, ,
\label{intrinsic}
\end{equation}
with
\begin{equation}
\tilde{q}_{\pi\nu}=u_{\nu}v_{\pi}\Sigma _{K}^{\nu\pi },\ \ \
q_{\pi\nu}=v_{\nu}u_{\pi}\Sigma _{K}^{\nu\pi},
\label{qs}
\end{equation}
in terms of the occupation amplitudes for neutrons and protons $v_{\nu,\pi}$   
($u^2_{\nu,\pi}=1-v^2_{\nu,\pi}$) and the matrix elements of the spin operator, 
$\Sigma _{K}^{\nu\pi}=\left\langle \nu\left| \sigma _{K}\right| \pi\right\rangle $, 
connecting proton and neutron single-particle states, 
as they come out from the HF+BCS calculation. 
$X_{\pi\nu}^{\omega _{K}}$ and 
$Y_{\pi\nu}^{\omega _{K}}$ are the forward and backward amplitudes of the 
pnQRPA phonon operator, respectively.

Once the intrinsic amplitudes in Eq. (\ref{intrinsic}) are calculated, 
the GT strength $B_{\omega}(GT^{-})$ in the laboratory system for a 
transition  $I_iK_i (0^+0) \rightarrow I_fK_f (1^+K)$ can be evaluated.
Using the Bohr-Mottelson factorization \cite{bm} to express the initial 
and final states in the laboratory system in terms of intrinsic states, 
we arrive at

\begin{eqnarray}
B_{\omega}(GT^{-} )& =& \sum_{\omega_{K}} \left[ \left\langle \omega_{K=0}
\left| \sigma_0t^{-} \right| 0 \right\rangle ^2 \delta (\omega_{K=0}-
\omega ) \right.  \nonumber  \\
&& \left. + 2 \left\langle \omega_{K=1} \left| \sigma_1t^{-} \right|
0 \right\rangle ^2 \delta (\omega_{K=1}-\omega ) \right] \, ,
\label{bgt}
\end{eqnarray}
in $[g_A^2/4\pi]$ units. 
The strength distributions will be relative to the excitation energy in the 
daughter nucleus, and are given by

\begin{equation}
E_{ex}=\omega_{\rm QRPA} -E_{\pi_0}-E_{\nu_0}\, ,
\end{equation}
where $E_{\pi_0}$ and $E_{\nu_0}$ are the lowest quasiparticle energies for
protons and neutrons, respectively.

The $\beta$-decay half-life is obtained by summing all the allowed
transition strengths to states in the daughter nucleus with
excitation energies lying below the corresponding $Q_\beta $-energy,

\begin{eqnarray}
Q_{\beta^-} &=& M(A,Z) - M(A,Z+1) - m_e \nonumber \\ 
&=&  BE(A,Z) - BE(A,Z+1) + m_n - m_p - m_e \, , 
\end{eqnarray}
written in terms of the nuclear masses $M(A,Z)$ or nuclear binding energies $BE(A,Z)$
and the neutron $(m_n)$, proton $(m_p)$, and electron mass $(m_e)$. 
The half-lives could be calculated in a quasiparticle approximation that avoids a 
direct calculation of $Q_{\beta}$ in terms of the parent and daughter masses 
\cite{engel}. This is achieved by expressing the binding energy of the daughter 
nucleus in terms of the binding energy of the parent, the Fermi energies for
protons and neutrons and the energy of the lowest two-quasiparticle excitation
$E_{\pi_0}+E_{\nu_0}$. In the present work we prefer to evaluate directly $Q_{\beta}$ 
from the binding energies of parent and daughter nuclei. This formulation is 
convenient to compare half-lives calculated from different $Q_{\beta}$ values 
that can be taken directly from experimental masses or from various existing 
mass formulas. Actually, the values obtained for $Q_{\beta}$ from binding energies 
and from the approximation in terms of Fermi levels and quasiparticle energies 
are very close to each other.

The weighting
coefficients are given by the phase space factors $f(Z,Q_{\beta}-E_{ex})$,

\begin{equation}
T_{1/2}^{-1}=\frac{\left( g_{A}/g_{V}\right) _{\rm eff} ^{2}}{D}
\sum_{0 < E_{ex} < Q_\beta}f\left( Z,Q_{\beta}-E_{ex} \right) B(GT,E_{ex}) \, ,
 \label{t12}
\end{equation}
with $D=6143$~s and $(g_A/g_V)_{\rm eff}=0.77(g_A/g_V)_{\rm free}$,
where 0.77 is a standard quenching factor and $(g_A/g_V)_{\rm free}=-1.270$. 
The same quenching factor is included in all the figures shown later for the 
GT strength distributions. The Fermi integral $f(Z,Q_{\beta}-E_{ex})$ is computed 
numerically for each value of the energy including screening and finite size 
effects, as explained in Ref. \cite{gove}.

In this work we consider only allowed GT transitions. It is well known that 
first-forbidden (FF) transitions gain relevance as $Q_{\beta}$ increases because 
the corresponding phase-space factors involve a quadratic dependence on the 
$\beta$-energy ($Q_{\beta}-E_{ex}$), which is absent in allowed GT transitions. As a 
result, the phase factors for allowed transitions scale as $(Q_{\beta}-E_{ex})^5$, 
whereas for forbidden transitions they scale as $(Q_{\beta}-E_{ex})^7$.
Consequently, one should take care of FF transitions as $Q_{\beta}$ becomes larger 
in very neutron-rich nuclei. Nevertheless, the calculations available for the 
FF strength in this mass region, based on different QRPA approaches 
\cite{marketin16,mustonen16}, show that although the relative contribution of 
FF transitions to the total rates increases with $Q_{\beta}$ as the number of 
neutrons increases, their contribution never exceeds an 8\% effect in 
Ref. \cite{mustonen16} and a 13\% effect in Ref. \cite{marketin16} for the 
heaviest isotopes of Ca, Ti, and Cr considered in this work. They are 
much smaller in lighter isotopes. 
Thus, according to those works, taking into account FF transitions will 
reduce the half-lives by 10\% at most in the heaviest nuclei considered 
in each isotopic chain and by a negligible amount in lighter isotopes.
This is not a relevant contribution for the purpose 
of this paper and can be safely neglected.

The $\beta$-feedings $I_\beta (E_{ex})$  (\%) are given by
\begin{equation}
 I_\beta (E_{ex}) (\% ) = 100 \frac{(g_A/g_V)_{\rm eff}^2}{D} f(Z,Q_\beta -E_{ex}) 
B(GT,E_ {ex})  T_{1/2}
\ .
\label{feed}
\end{equation}

The probability of $\beta$-delayed neutron emission is given by

\begin{equation}
P_n = \frac{ {\displaystyle \sum_{S_n < E_{ex} < Q_\beta}f\left( Z,Q_{\beta}-E_{ex}
\right) B(GT,E_{ex}) }}
{{\displaystyle \sum_{0 < E_{ex} < Q_\beta}f\left( Z,Q_{\beta}-E_{ex} \right)
B(GT,E_{ex})}}\, ,
\label{pn}
\end{equation}
where the sums extend to all the excitation energies in the daughter nuclei 
in the indicated ranges. $S_n$ is the one-neutron separation energy in the 
daughter nucleus. According to Eq.~(\ref{pn}), $P_{n}$ is mostly sensitive 
to the strength located at energies around $S_n$, thus providing a structural
probe complementary to $T_{1/2}$.
Eq. (\ref{pn}) assumes that all the decays to excited states in the daughter 
nucleus with energies above $S_n$ always lead to delayed neutron emission and 
then, $\gamma$-decay from neutron unbound levels is neglected. Recent studies 
coupling the microscopic QRPA and the statistical Hauser-Feshbach model show 
that the competition between neutron and $\gamma$ emission can modify the 
neutron emission probabilities in a way that depends on the system considered, 
but is enhanced when approaching the neutron drip-lines \cite{kawano08,mumpower16}. 

\section{Results and discussion}
\label{results}

We first show the results obtained for the PECs in the isotopes studied. 
The energy distribution of the GT strength corresponding to the local
minima of the PECs is studied afterwards. Finally, half-lives and 
$\beta$-delayed neutron emission probabilities are computed.

\begin{figure}[htb]
\centering
\includegraphics[width=70mm]{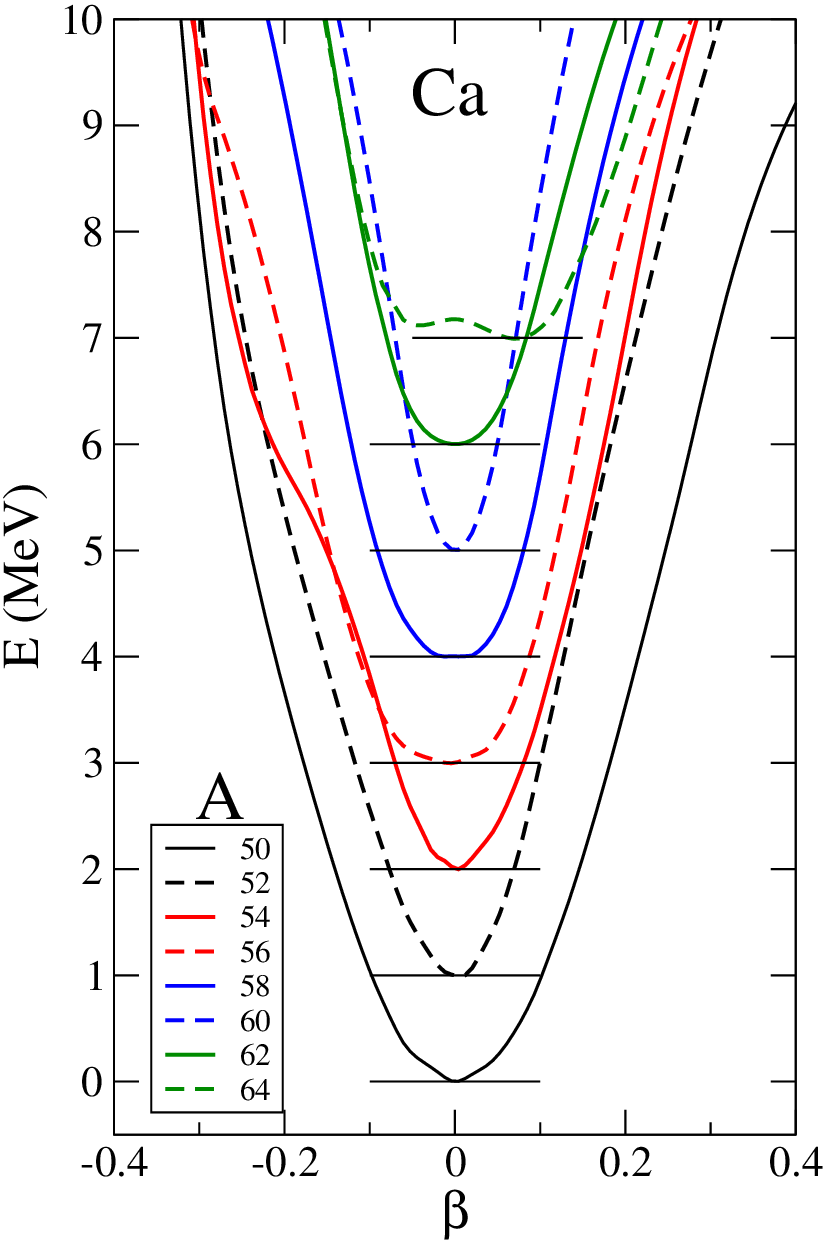}
\caption{Energy curves for $^{50,52,54,56,58,60,62,64}$Ca isotopes obtained from 
constrained HF+BCS calculations with the Skyrme force SLy4.
}
\label{fig_eq_ca}
\end{figure}

\begin{figure}[htb]
\centering
\includegraphics[width=70mm]{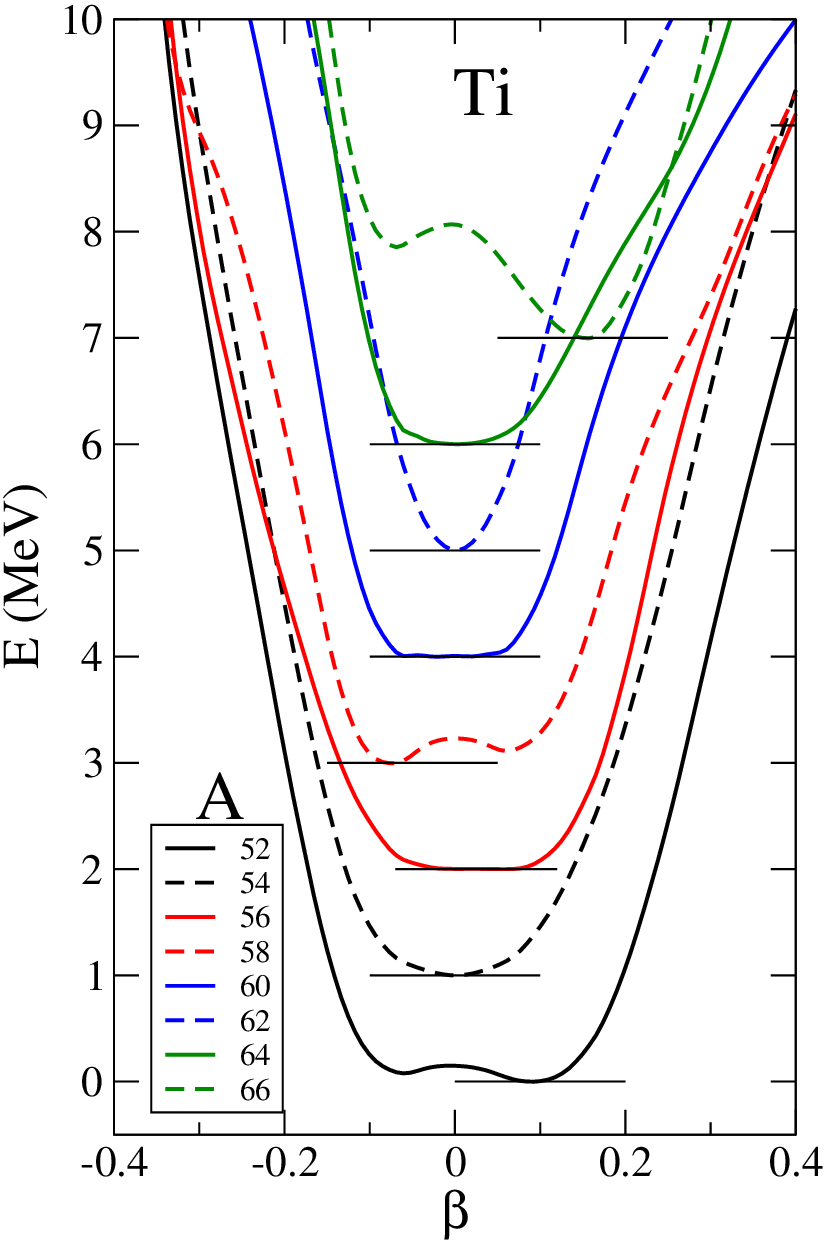}
\caption{Same as in Fig. \ref{fig_eq_ca}, but for $^{52,54,56,58,60,62,64,66}$Ti isotopes.
}
\label{fig_eq_ti}
\end{figure}

\begin{figure}[htb]
\centering
\includegraphics[width=70mm]{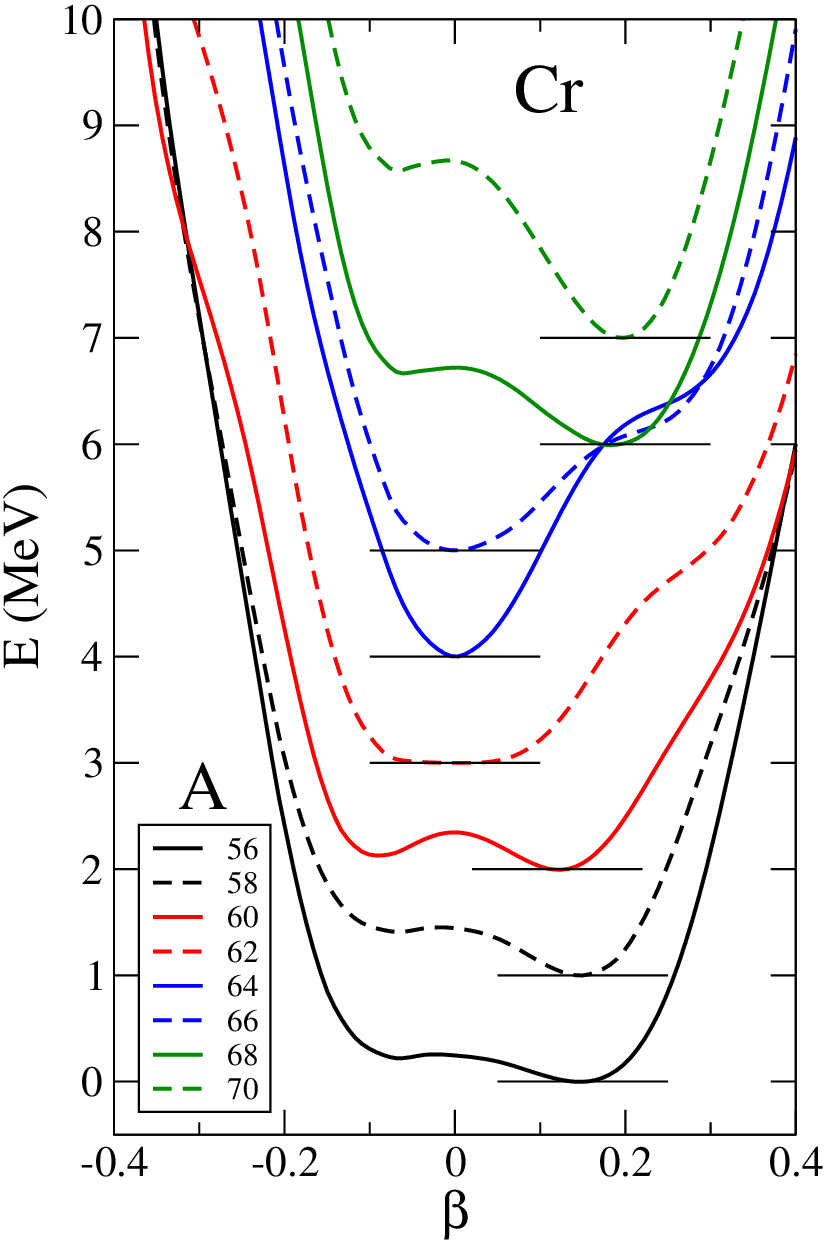}
\caption{Same as in Fig. \ref{fig_eq_ca}, but for $^{56,58,60,62,64,66,68,70}$Cr isotopes.
}
\label{fig_eq_cr}
\end{figure}

\subsection{Potential Energy Curves}

In Figs. (\ref{fig_eq_ca} - \ref{fig_eq_cr}), we show the PECs obtained from 
SLy4, relative to the ground state energy, as a function of the quadrupole 
deformation $\beta$ for neutron-rich Ca ($Z=20$), Ti ($Z=22$), and Cr ($Z=24$), 
respectively. Neutron numbers cover the range $N=30-46$. For a better comparison, 
the plot corresponding to each isotope has been shifted by 1 MeV relative to the 
neighbor lighter isotope. 

In Fig. \ref{fig_eq_ca} we observe spherical solutions in practically all 
semi-magic Ca isotopes. Only the heavier $^{64}$Ca with the last neutrons occupying 
partially the $1g_{9/2}$ orbital exhibits deformed solutions. We also observe that  
$^{52}$Ca ($N=32$) and $^{60}$Ca ($N=40$) show relatively sharper profiles, as they 
correspond to the extra stability provided by the $2p_{3/2}$ and $1f_{5/2}-2p_{1/2}$ 
sub-shell closures, respectively.
Fig. \ref{fig_eq_ti} for Ti isotopes shows a more involved structure of the PECs. 
Similarly to the above case of Ca isotopes, we observe a tendency toward spherical
shapes in $^{54}$Ti ($N=32$) and  $^{62}$Ti ($N=40$), related to the same sub-shell
closures mentioned above, but in general the PECs are not so sharp as for Ca 
isotopes and deformed shapes, oblate and prolate, are developed between sub-shell 
closures. Fig. \ref{fig_eq_cr} for Cr isotopes shows in most cases prolate 
structures that become more pronounced as we move away from  $^{64}$Cr where the 
sub-shell closure at $N=40$ is still manifest. In some cases a second minima in 
the oblate sector appears, although it is very shallow and not well separated 
from the prolate minimum by energy barriers.

\begin{figure}[htb]
\centering
\includegraphics[width=70mm]{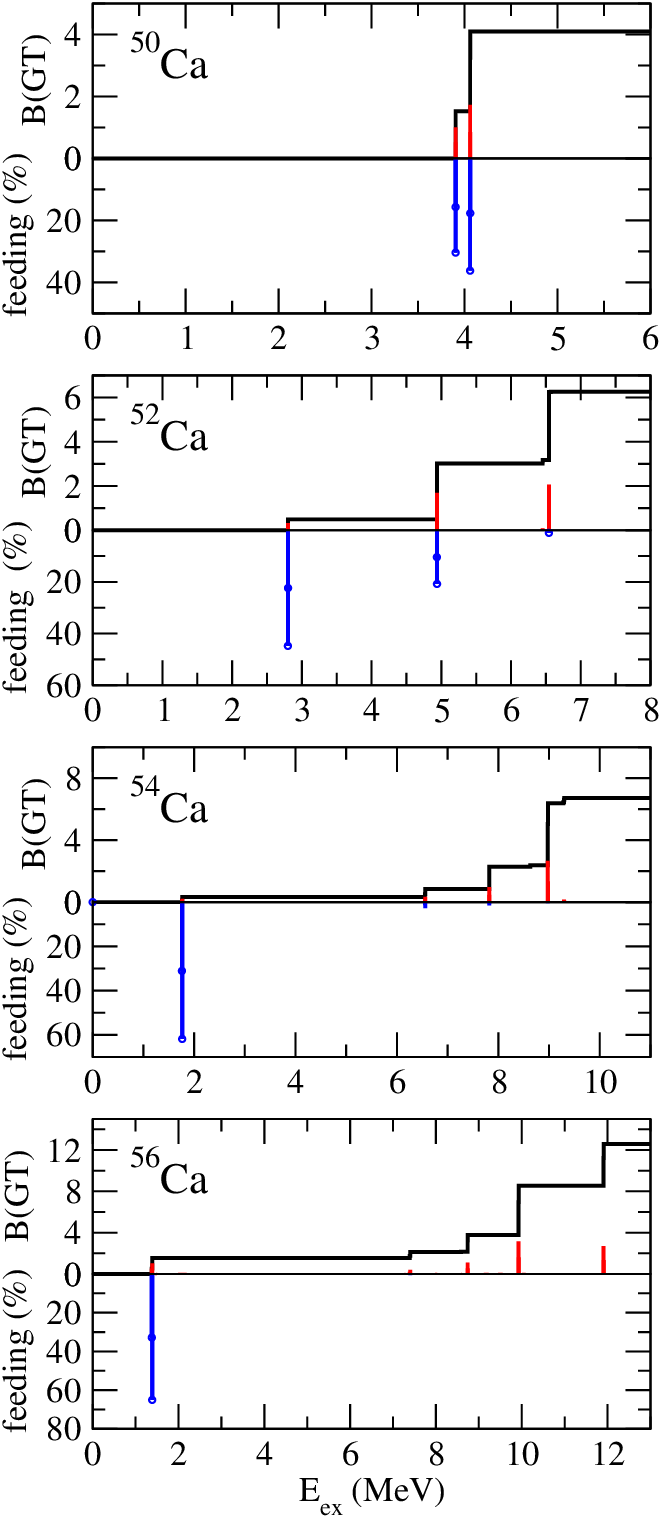}
\caption{Gamow-Teller strength distributions, $B(GT)$, cumulative sum strength, and 
$\beta$-feedings for $^{50,52,54,56}$Ca isotopes.
}
\label{fig_ca_1_bgt}
\end{figure}

\begin{figure}[htb]
\centering
\includegraphics[width=70mm]{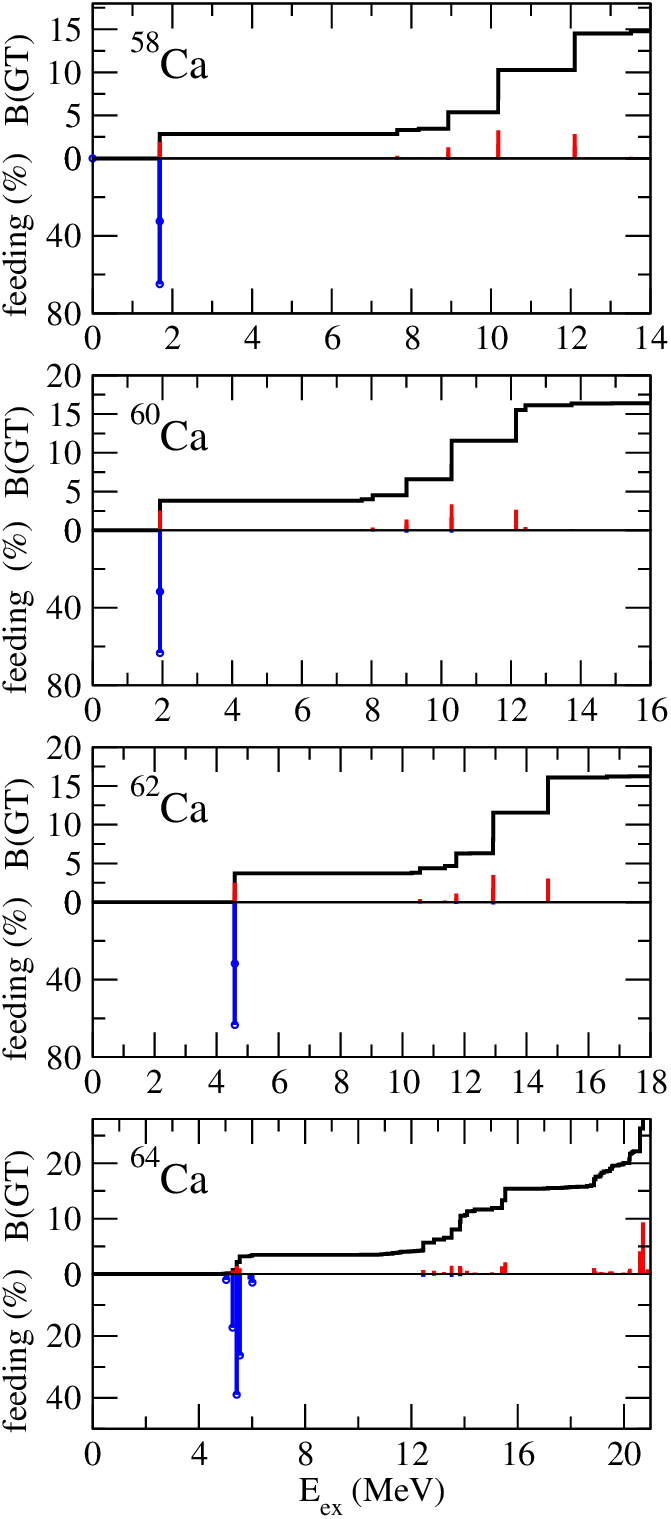}
\caption{Same as in Fig. \ref{fig_ca_1_bgt}, but for $^{58,60,62,64}$Ca isotopes.
}
\label{fig_ca_2_bgt}
\end{figure}

\begin{figure}[htb]
\centering
\includegraphics[width=70mm]{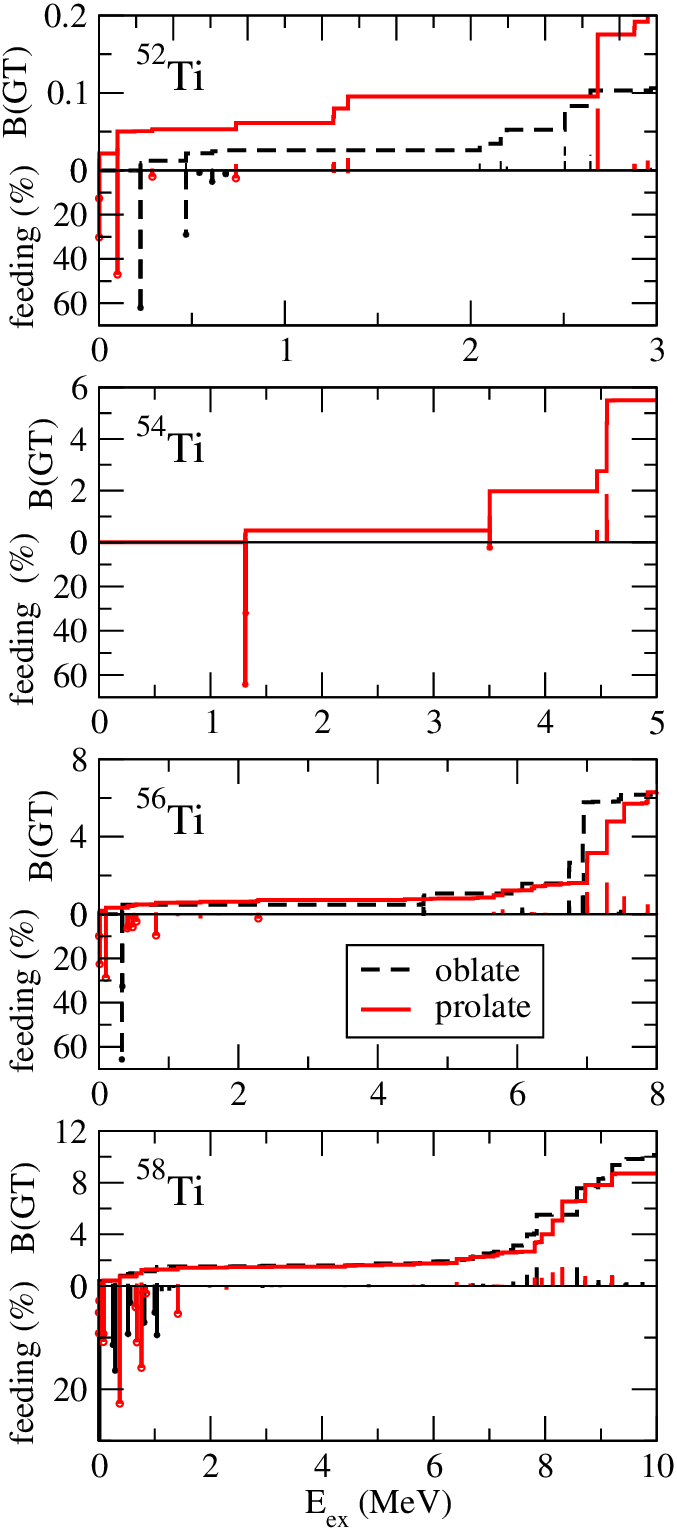}
\caption{Same as in Fig. \ref{fig_ca_1_bgt}, but for $^{52,54,56,58}$Ti isotopes.
Results obtained from oblate and prolate shapes are shown.}
\label{fig_ti_1_bgt}
\end{figure}

\begin{figure}[htb]
\centering
\includegraphics[width=70mm]{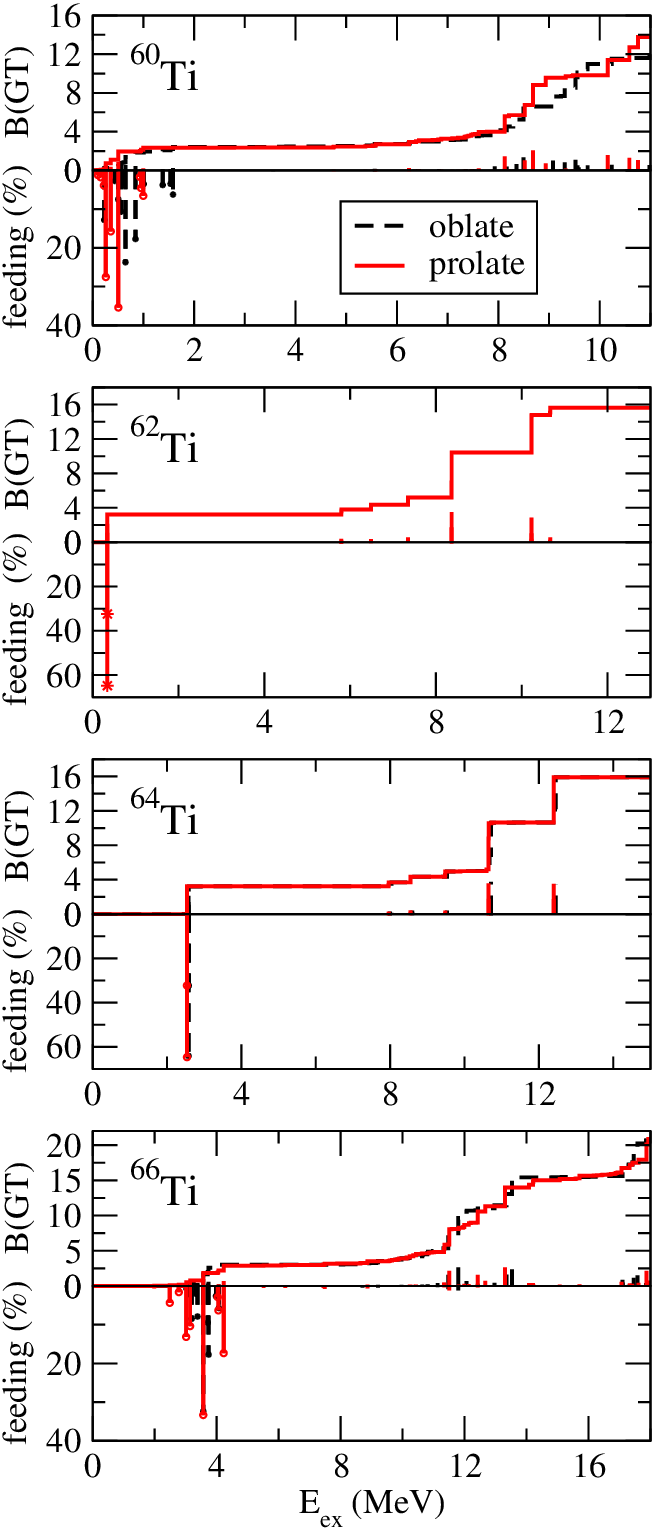}
\caption{Same as in Fig.  \ref{fig_ti_1_bgt}, but for $^{60,62,64,66}$Ti isotopes.
}
\label{fig_ti_2_bgt}
\end{figure}

\begin{figure}[htb]
\centering
\includegraphics[width=70mm]{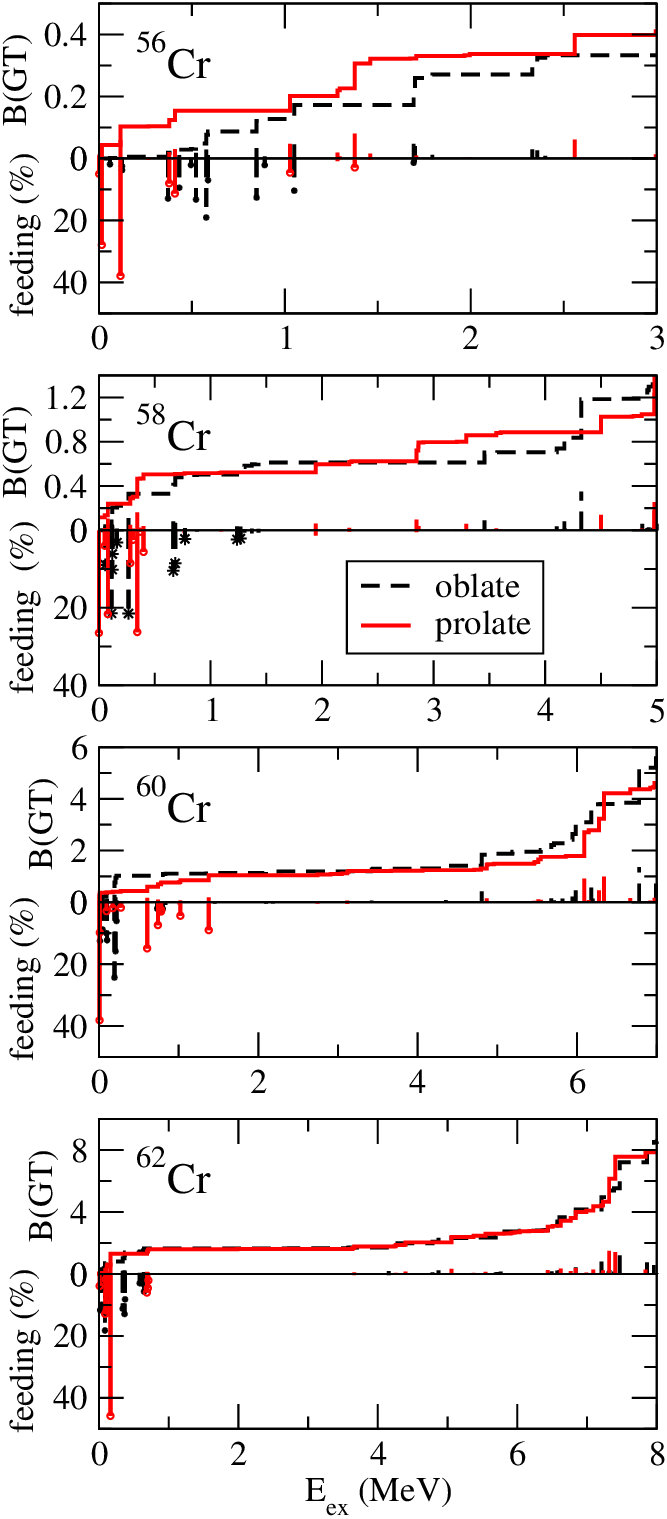}
\caption{Same as in Fig. \ref{fig_ti_1_bgt}, but for $^{56,58,60,62}$Cr isotopes.
}
\label{fig_cr_1_bgt}
\end{figure}

\begin{figure}[htb]
\centering
\includegraphics[width=70mm]{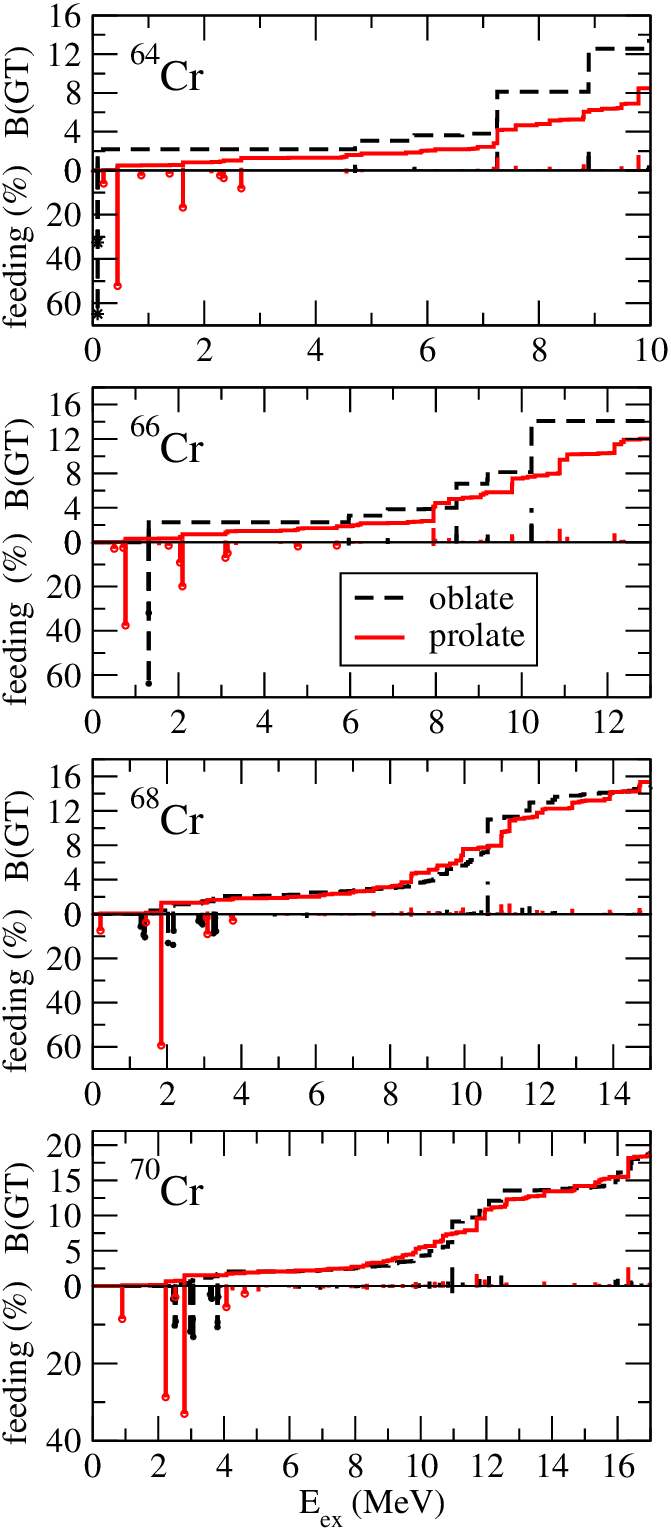}
\caption{Same as in Fig. \ref{fig_ti_1_bgt}, but for $^{64,66,68,70}$Cr isotopes.
}
\label{fig_cr_2_bgt}
\end{figure}

\begin{figure}[htb]
\centering
\includegraphics[width=70mm]{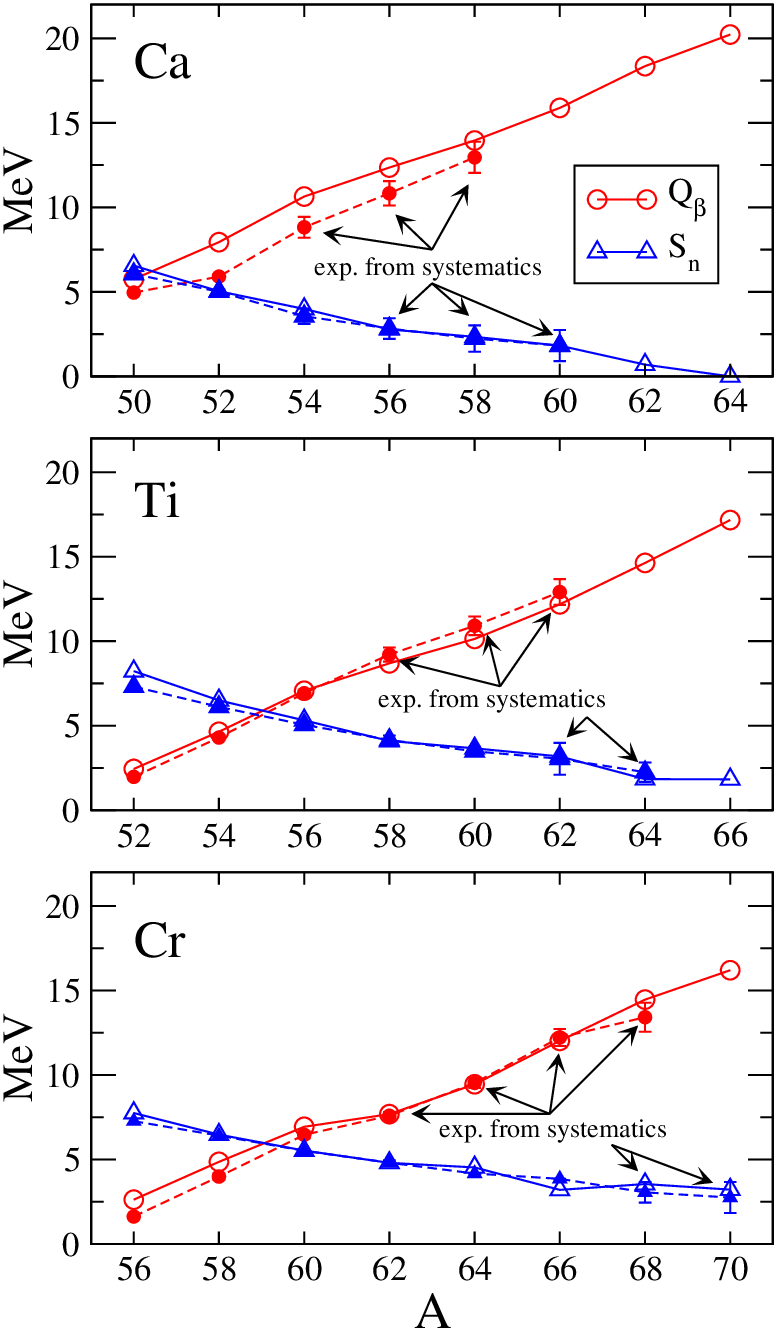}
\caption{$Q_\beta$ energies (MeV) for Ca, Ti, and Cr isotopes 
and $S_n$ energies (MeV) in the corresponding daughter nuclei,
as a function
of the mass number $A$. Open symbols are the results from SLy4-pnQRPA calculations,
whereas solid symbols with error bars stand for the experimental data \cite{audi12}. 
The experimental data points of the heavier isotopes are obtained from systematics
\cite{audi12}.
}
\label{fig_sn_qb}
\end{figure}

\subsection{Gamow-Teller strength distributions}

The energy distribution of the GT strength is quite important to constrain 
the underlying nuclear structure. For a theoretical model, it represents 
a more demanding test than just reproducing half-lives that are integral 
quantities obtained from these strength 
distributions properly weighted with phase factors (see Eq. (\ref{t12})). 
This is of especial importance in astrophysical scenarios of high 
densities and temperatures that cannot be reproduced in the laboratory.
Given that the phase factors in the stellar medium are different from 
those in the laboratory, to describe properly the decay rates under extreme 
conditions of density and temperature, it is not sufficient to reproduce the
half-lives in the laboratory. One needs, in addition, to have a reliable 
description of the GT strength distributions.

In Figs. (\ref{fig_ca_1_bgt}-\ref{fig_cr_2_bgt}), we show in the upper plots 
of each panel the results for the energy distributions of the quenched GT 
strength (BGT). We can see both the individual GT strengths and the cumulative
sums that give the total strength accumulated up to a given energy.
In the lower plots we show the $\beta$-feedings from Eq. (\ref{feed}). They are
plotted versus the excitation energy of the daughter nucleus.
The results correspond to the (oblate-prolate-spherical) equilibrium shapes 
for which we obtained minima in the PECs.
The interval of excitation energies for each isotope is limited by the $Q_\beta$ 
energy, which is the relevant energy range for the calculation of the decay 
properties. 

In Figs. \ref{fig_ca_1_bgt} and \ref{fig_ca_2_bgt} we show the results for Ca 
isotopes. In this case, according to the PECs in Fig. \ref{fig_eq_ca}, 
calculations from one single shape (spherical) are performed.
Although the GT strengths are calculated within pnQRPA, it is worth analyzing
their distributions in terms of quasiparticle transitions that offer a simple 
and natural tool to interpret the underlying excitation mechanism. In the case
of Ca isotopes this analysis can be made in terms of the spherical orbitals.
In allowed transitions parity is conserved and therefore, only orbitals in the
$fp$-shell are connected by the GT operator, except in the heavier isotopes
where transitions within the $1g_{9/2}$ shell start to contribute as well. 
Thus, the low-lying GT strength in all Ca isotopes studied correspond to the
transition  $\nu 1f_{5/2} \rightarrow \pi 1f_{7/2}$. The excitation energy of this
transition is stabilized in the vicinity of 2 MeV, except for the lighter
$^{50,52}$Ca and the heavier $^{62,64}$Ca isotopes, where the excitation occurs
at somewhat higher energies. These transitions account for most of the feeding 
in the decay.
After a relatively large energy gap without strength, we find next in energy 
transitions corresponding to $\nu 1f_{7/2} \rightarrow\pi 1f_{7/2}$, 
$\nu 2p_{3/2} \rightarrow \pi 2p_{3/2}$, and $\nu 2p_{1/2} \rightarrow\pi 2p_{3/2}$. 
These transitions are responsible for the structure of peaks observed in $^{54-58}$Ca,
as we populate with neutrons the $2p_{1/2}$ and $1f_{5/2}$ orbitals. The GT strength
patterns of these isotopes are very similar to each other.

Other competing transitions between negative parity partners also appear either 
at higher energies or having smaller strength. In the heavier isotopes $^{62,64}$Ca, 
transitions between positive
parity states $\nu 1g_{9/2} \rightarrow \pi 1g_{g/2}$  appear beyond 10 MeV.
In the case of $^{64}$Ca, the GT strength appears more fragmented because of the
effect of deformation that splits energetically the spherical orbitals. In this
case the results correspond to the prolate minimum in the PEC of Fig. \ref{fig_eq_ca}.

In the cases of Ti and Cr isotopes (Figs. \ref{fig_ti_1_bgt} - \ref{fig_cr_2_bgt}), 
the analysis can be done similarly, but the interpretation is more involved
because of the large fragmentation caused by deformation.
The PECs of Ti isotopes in Fig. \ref{fig_eq_ti} show that $^{54}$Ti, $^{62}$Ti, and
$^{64}$Ti are spherical and the corresponding GT strength distributions show
profiles characterized by isolated peaks similar to those for Ca isotopes discussed
above. Actually, the analysis of the structure of these transitions shows that
the low-lying excitations correspond again to  $\nu 1f_{5/2} \rightarrow \pi 1f_{7/2}$
and similarly for the analysis of higher excitations. In the case of deformed shapes,
the strength is clearly more fragmented and the analysis should be done in 
terms of Nilsson states. Cr isotopes, in figures \ref{fig_cr_1_bgt} and
\ref{fig_cr_2_bgt}, show clear deformed patterns, except in the case of
$^{54}$Cr with $N=40$ related to the $1f_{5/2}-2p_{1/2}$ sub-shell closure.

The general structure observed in the profiles of the strength distributions 
in this mass region is then characterized by some strength at very low excitation 
energy that, although not very large, carries most of the feeding and is very 
significant to determine the half-lives.
Then, there is a large energy region of several MeV with practically no strength,
and a higher energy region where the strength concentrates.

The calculations the GT strength distributions for the various equilibrium shapes
show us the sensitivity of these observables to the nuclear shape. It has
been shown in the past that in particular cases, a strong sensitivity of 
the distribution profiles to the deformation is apparent. This feature was 
exploited to gain information about the nuclear shape of the decaying
nucleus \cite{exppoirier,expnacher,expperez,expestevez,expbriz}.
In the present case, the effect of the nuclear shape on the GT strength
distributions is of minor importance, as shown in Figs. 
\ref{fig_ti_1_bgt} - \ref{fig_cr_2_bgt}.

\begin{figure}[htb]
\centering
\includegraphics[width=60mm]{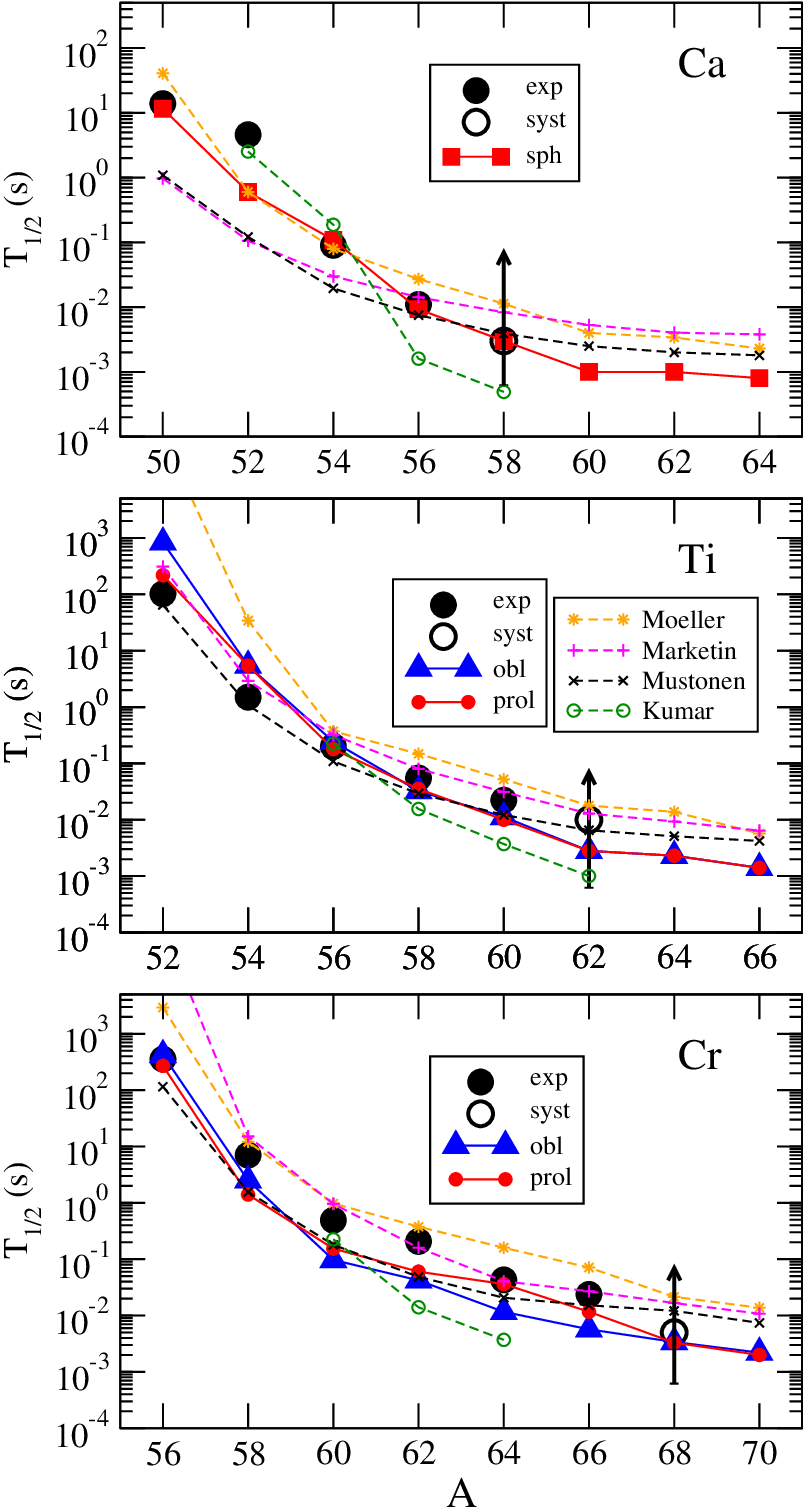}
\caption{$\beta$-decay half-lives, $T_{1/2}$ (s), for Ca, Ti and Cr isotopes.
Our results are compared with experiment \cite{audi12} and other theoretical
calculations (see text).
}
\label{fig_t12}
\end{figure}

\begin{figure}[htb]
\centering
\includegraphics[width=60mm]{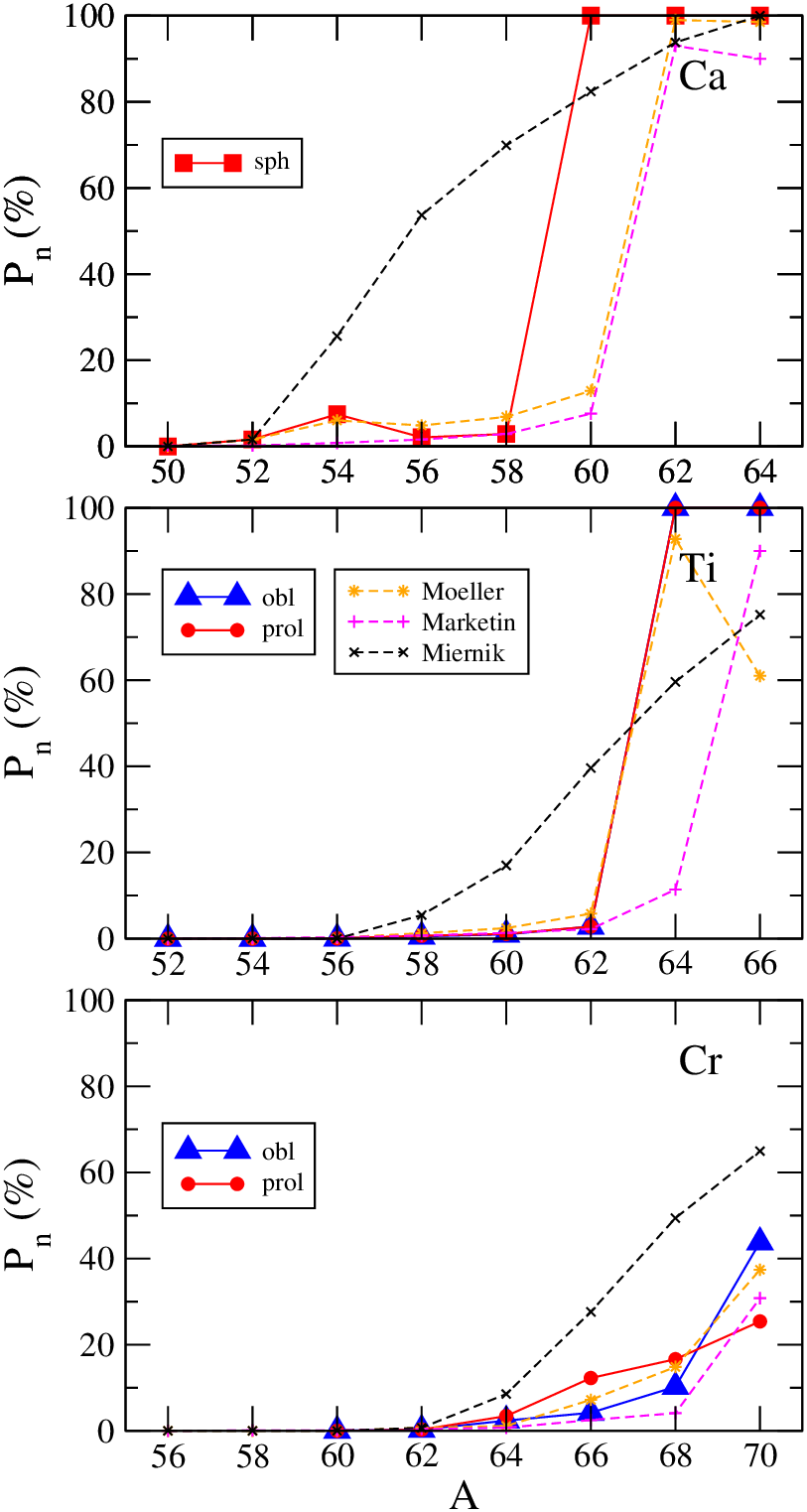}
\caption{$\beta$-delayed neutron-emission probabilities $P_n\ (\%)$.
Our results are compared with theoretical results from Refs. 
\cite{moller3,marketin16,miernik13}.
}
\label{fig_pn}
\end{figure}

\subsection{Half-lives and $\beta$-delayed neutron-emission probabilities}

In this section we present the results for $T_{1/2}$ and $P_n$. 
The calculation of the half-lives in Eq. (\ref{t12}) involves knowledge
of the GT strength distribution as well as of the $Q_{\beta}$ energies.
The calculation of $P_n$ requires in addition knowledge of the one-neutron 
separation energies in the daughter nuclei $S_n$. 
Experimental values of these quantities based on nuclear
masses are available for some of the isotopes studied, but not for the most
exotic. Then, in order to present a unified description of all the isotopes,
we use in this work $Q_{\beta}$ and $S_n$ energies evaluated from theoretical
calculations of the masses, based on the SLy4 Skyrme force.

Fig. \ref{fig_sn_qb} shows a comparison between calculated and experimental
$Q_{\beta}$ and  $S_n$ energies. Experimental data with error bars are taken 
from Ref. \cite{audi12}
and plotted with solid symbols, circles for $Q_{\beta}$ and triangles for $S_n$.
Arrows point to the energies that are not directly measured, but are taken
from systematics. The calculated energies are plotted with similar open symbols.
The agreement with experiment is quite satisfactory, except in the $Q_{\beta}$ 
values of the lighter Ca isotopes, where the data are overestimated.

In Fig. \ref{fig_t12} we show our results for the $\beta$-decay half-lives 
in the Ca, Ti, and Cr isotopic chains. The results are compared with experiment
\cite{audi12} (open circles are taken from systematics).
We also compare our results with other theoretical calculations. 
The results labeled as 'Moeller' \cite{moller3} include contributions from GT 
and first-forbidden transitions. While the former correspond to microscopic 
pnQRPA calculations using a Yukawa single-particle Hamiltonian and a separable
residual interaction in the ph channel, the latter are obtained
from a statistical gross theory. The $Q_{\beta}$ and $S_n$ energies are 
evaluated from the masses calculated in the finite-range droplet model.
The calculations are done without any quenching of the axial-vector coupling 
constant $g_A$.
The results labeled as 'Marketin' \cite{marketin16} correspond to a relativistic 
formalism that includes FF transitions with masses calculated from the same model. 
Spherical symmetry is assumed in these calculations and quenched  
$g_A$ values are used in both GT and FF.
The results labeled as 'Mustonen' \cite{mustonen16} are based on
Skyrme pnQRPA for axially deformed even-even nuclei. Masses are calculated
consistently with the same interaction and quenching of $g_A$ is only included
for GT transitions.
Finally, the half-lives labeled as 'Kumar' \cite{kumar16} correspond to shell-model 
results.

The agreement of our results with experiment is in general satisfactory. The 
general trends are well reproduced, although discrepancies are found in 
particular cases. This is the case of $^{52}$Ca, where we underestimate the 
half-life. In the case of Ti isotopes the agreement is fairly good, although 
the lighter isotopes $^{52,54}$Ti are overestimated. For Cr isotopes we 
underestimate the data, especially in $^{58,60,62}$Cr isotopes.
Results from oblate and prolate shapes in Ti and Cr isotopes can be also
observed in the figure. In most cases, prolate configurations seem to describe
the data somewhat better and this sensitivity could be exploited experimentally
to discriminate in favor of one of the shapes.

The agreement of our results with experiment is comparable to the agreement
obtained with the other calculations from different theoretical formalisms.
The results from Ref. \cite{moller3} show a general tendency to overestimate
the half-lives, whereas those of Ref. \cite{kumar16} tend to underestimate
the data. In Ca isotopes the calculations from both calculations 
\cite{marketin16} and \cite{mustonen16} clearly underestimate the data in 
the lighter isotopes. The agreement with Ti isotopes is remarkable in both 
calculations. In the case of Cr isotopes, results from Ref. \cite{marketin16}  
(\cite{mustonen16}) tend to overestimate (underestimate) the experiment. 

Fig. \ref{fig_pn} depicts the results for $P_n$. Our results for the various
shapes are compared with results from Refs. \cite{moller3,marketin16}, as
well as with the results from Miernik \cite{miernik13} that correspond to a 
phenomenological model based on a statistical level density function with 
masses obtained from the Skyrme interaction HFB-21.
Unfortunately, there are no experimental data available yet for these isotopes,
but measurements of $P_n$ values and half-lives are under consideration
in this region within the BRIKEN Collaboration \cite{dillmann18} at RIKEN.

Our results, as well as the other microscopic calculations, show a sudden 
shift from almost zero values of $P_n$ to practically 100 \% in Ca and Ti 
isotopes. This is related to fact that the lack of fragmentation in the
GT strength distribution makes the energy $S_n$ very critical to determine 
whether or not all the strength is contained beyond $S_n$.
Thus, in $^{58}$Ca the energy $S_n=2.34$ MeV is above the energy 1.68 MeV that
receives practically all the feeding, while not generating neutron emission.
On the other hand, in $^{60}$Ca the energy $S_n=1.82$ MeV is already below
the energy 1.92 MeV that receives the feeding. Then, almost all the
feeding in the decay is received by unbound states.
Similarly in the case of Ti isotopes, $P_n$ is almost zero in  $^{62}$Ti
because $S_n=3.18$ MeV is clearly above the excitation energy 0.34 MeV 
that takes all the feeding, whereas in $^{64}$Ti, $S_n=1.84$ MeV is below the 
excitation energy 2.53 MeV that receives the most of the feeding. 
This effect does not happen in deformed nuclei, where the fragmentation of 
the strength makes the exact position of $S_n$ not so critical,
and consequently $P_n$ increases smoothly. This is the case of the Cr isotopes 
where the fragmented strength induced by deformation is translated into a more 
continuous increase of $P_n$, causing a sensitivity to the nuclear shape
that could be exploited experimentally.

\section{Summary and Conclusions}

$\beta$-decay properties including energy distributions of the GT strength 
and feedings, half-lives, and delayed neutron emission are studied in 
neutron-rich even-even Ca, Ti, and Cr isotopic chains. This mass region
is important for the isotopic evolution of nuclear structural effects. Nuclei
in this mass region are also implicated in the astrophysical URCA process,  
a neutrino cooling mechanism at work in white dwarfs and neutron stars.

The theoretical approach is based on a pnQRPA on top of a deformed HF+BCS
with Skyrme forces. Residual forces in the ph and pp channels are included
as well. We first calculate the GT strength distributions that result
from the underlying nuclear structure and from which other integral 
quantities are evaluated afterwards.
Unfortunately, there are still no data on strength distributions to compare 
with, and these calculations are for the moment predictions. The results are 
compared with the experimental information available on half-lives. In general
we get a fair agreement, which is comparable with the agreement achieved
with other calculations using different theoretical approaches. 
$\beta$-delayed neutron emission probabilities are also calculated and 
compared with the results from other models.
These calculations are timely because of the increased capabilities of modern 
existing facilities (RIKEN,NCSL-MSU) and new setups to reach more exotic nuclei 
and measure these quantities. The results presented in this work will also allow 
further testing of different theoretical models for calculations of the decay 
properties of exotic nuclei of astrophysical interest. 

\begin{acknowledgments}
This work was supported by Ministerio de Econom\'\i a y Competitividad
(Spain) under Contracts FIS2014-51971-P, FPA2014-52823-C2-1-P, and the Severo 
Ochoa program (SEV-2014-0398).
\end{acknowledgments}

\end{document}